\documentclass[%
 reprint,
 superscriptaddress,
 amsmath,amssymb,
 aps,
]{revtex4-1}

\usepackage{graphicx}
\usepackage{dcolumn}
\usepackage{bm}
\usepackage{diagbox}
\usepackage{adjustbox}   
\usepackage[graphicx]{realboxes}

\usepackage[utf8]{inputenc}
\usepackage[T1]{fontenc}
\usepackage{xcolor}

\begin{document}

\title{Hierarchical clustering of bipartite data sets\\ based on the statistical significance of coincidences}

\author{Ignacio Tamarit}
\affiliation{%
 Grupo Interdisciplinar de Sistemas Complejos (GISC),
}%
\affiliation{%
 Departamento de Matem\'aticas de la Universidad Carlos III de Madrid, Legan\'es, Spain
}%
\affiliation{%
 Unidad Mixta Interdisciplinar de Comportamiento y Complejidad Social (UMICCS), Madrid, Spain
}%


\author{Mar\'{\i}a Pereda}
\affiliation{%
 Grupo Interdisciplinar de Sistemas Complejos (GISC),
}%
\affiliation{%
 Unidad Mixta Interdisciplinar de Comportamiento y Complejidad Social (UMICCS), Madrid, Spain
}%
\affiliation{%
 Grupo de Investigaci\'on Ingenier\'ia de Organizaci\'on y Log\'istica (IOL),
 Escuela T\'ecnica Superior de Ingenieros Industriales, Universidad Polit\'ecnica de Madrid, Madrid, Spain
}%

\author{Jos\'e A. Cuesta}
 \homepage{http://gisc.uc3m.es/~cuesta}
 \email{cuesta@math.uc3m.es}
\affiliation{%
 Grupo Interdisciplinar de Sistemas Complejos (GISC),
}%
\affiliation{%
 Departamento de Matem\'aticas de la Universidad Carlos III de Madrid, Legan\'es, Spain
}%
\affiliation{%
 Unidad Mixta Interdisciplinar de Comportamiento y Complejidad Social (UMICCS), Madrid, Spain
}%
\affiliation{%
 Instituto de Biocomputaci\'on y F\'isica de Sistemas Complejos (BIFI),
Universidad de Zaragoza, 
Zaragoza, Spain
}%
\affiliation{%
 UC3M-Santander Big Data Institute (IBiDat), 
Getafe, Spain
}%

\date{\today}

\begin{abstract}
When some `entities' are related by the `features' they share they are amenable to a bipartite network representation. Plant-pollinator ecological communities, co-authorship of scientific papers, customers and purchases, or answers in a poll, are but a few examples. Analyzing clustering of such entities in the network is a useful tool with applications in many fields, like internet technology, recommender systems, or detection of diseases. The algorithms most widely applied to find clusters in bipartite networks are variants of modularity optimization. Here we provide an hierarchical clustering algorithm based on a dissimilarity between entities that quantifies the probability that the features shared by two entities is due to mere chance. The algorithm performance is $O(n^{2})$ when applied to a set of $n$ entities, and its outcome is a dendrogram exhibiting the connections of those entities. Through the introduction of a `susceptibility' measure we can provide an `optimal' choice for the clustering as well as quantify its quality. The dendrogram reveals further useful structural information though---like the existence of sub-clusters within clusters or of nodes that do not fit in any cluster. We illustrate the algorithm by applying it first to a set of synthetic networks, and then to a selection of examples. We also illustrate how to transform our algorithm into a valid alternative for one-mode networks as well, and show that it performs at least as well as the standard, modularity-based algorithms---with a higher numerical performance. We provide an implementation of the algorithm in Python freely accessible from GitHub.
\end{abstract}

\maketitle


\section{\label{sec:intro}Introduction}

Among the networks that we can find in real life, bipartite networks stand on their own because of their special nature. Bipartite (two-mode) networks divide their nodes into two different categories, and links join nodes of one category \emph{only} with nodes of the other. Bipartite networks can be used to describe plant-pollinator mutualistic interactions \cite{memmott:1999,bascompte:2003,dormann:2014}, words in documents \cite{srivastava:2013,dhillon:2001}, scientists and co-authored papers \cite{newman:2001a,newman:2001b}, genes in viral genomes \cite{iranzo:2016,shapiro:2018}, actors in films \cite{peixoto:2013,peixoto:2014}, people attending events \cite{freeman:2003}, recommender systems \cite{ramasco:2004,shang:2010,guimera:2012}, etc., and they have been successfully applied to problems ranging from internet technology  \cite{cai:2013,xu:2014} to systems biology and medicine \cite{pavlopoulos:2018}. A defining feature of any system amenable to bipartite-network modeling is that one set can be thought of as `entities' and the other one as `features'. For instance, if the entities are scientists, the features are papers they author---or vice versa, if the entities are the papers, the features are their authors. Which is the set of entities and which the set of features very much depends on the problem one aims to solve, because a typical question regarding this kind of datasets is: how do entities cluster according to their set of features?

Finding clusters (also called modules or communities) in networks has been an active topic of research for a few decades (see~\cite{fortunato:2016} and references therein). There is no clear-cut definition of what a cluster or community is. Intuitively, one expects that nodes in a cluster are more densely connected to each other than to the nodes outside the cluster, but the actual definition is part of the answer to the clustering problem. For this reason, there is a plethora of different methods to determine clusters, and although there is some overlapping in their outcomes, they hardly obtain exactly the same partition of the set of nodes. Which method to choose is then a problem-dependent issue.

Bipartite networks require a purposely definition, because of their very particular connectivity---nodes of the same type are never connected to each other. Roughly speaking, three kinds of approaches have been explored. The most direct one amounts to projecting the network on the type of nodes whose clustering is sought \cite{zhou:2007,araujo:2011,tumminello:2011}. The result is a weighted network linking these nodes and only these---to which standard clustering algorithms can be applied. The success of this approach very much relies on a suitable choice of the weights for the links.

The second approach is, so to speak, global in nature. A typical method amounts to defining a function of the partition of nodes (`modularity'), and then finding the partition that maximizes it \cite{newman:2004}. This function compares the actual linking of the network with what a random null model would produce. The clustering problem then becomes the problem of finding the partition that maximizes the modularity of the network. Extending this method to bipartite networks requires choosing a suitable null model \cite{barber:2007,guimera:2007}. An alternative to modularity is to adopt a Bayesian viewpoint by introducing a stochastic block model \cite{guimera:2012} whose parameters are obtained through likelihood maximization \cite{larremore:2014}. Although the use of these global methods to determine the community structure of a network is widespread, their application to large datasets is limited because they are computationally demanding (they boil down to performing a combinatorial optimization). Furthermore, the very definition of modularity has some resolution limitations that preclude these methods from detecting clusters that are particularly small \cite{fortunato:2007,radicchi:2010}.

The last approach to the problem is represented by a set of methods that go under the common name of \emph{hierarchical clustering} \cite[ch.~4]{everitt:2011}. The idea of hierarchical clustering is to define a `dissimilarity' (often a true mathematical `distance') between entities based on the features that they do or do not share, and then sequentially merge the least dissimilar clusters (initially every node is a cluster), following some prescription. The outcome of these methods is not a partition, but a \emph{dendrogram,} i.e., a rooted tree in which nodes are grouped according to the dissimilarity value at which they merged into the same cluster. They look very much like phylogenetic trees and can be interpreted similarly. If needed, one can obtain a partition out of a dendrogram either by introducing a dissimilarity threshold or by detecting groups of branches that separate very near the root. As a matter of fact, the seminal work on community detection in networks uses a particular form of hierarchical clustering \cite{girvan:2002}.

There are two main reasons why there is a current preference for global methods over hierarchical clustering. One is the fact that on the latter the definition of clusters eventually depends on the choice of an arbitrary threshold---or a similar \emph{ad hoc} criterion. The other is the vast amount of different dissimilarity measures that people have used in the literature \cite{gower:1986}---each one yielding a different result \cite[ch.~3]{everitt:2011}.  Nevertheless, the upside of these methods is that they can be computationally more efficient because they do not involve any combinatorial optimization process. If $n$ denotes the number of entities, hierarchical clustering algorithms exist with time complexity $O(n^2)$ \cite{sibson:1973,defays:1977}. If one is willing to sacrifice exactness and can cope with approximate results, algorithms can be found that reduce this complexity to $O(n\log n)$ \cite{xie:2020} or even $O(n)$ using hash tables \cite{koga:2007}. (Notice, however, that the outcome of these algorithms depends on the right tuning of a bunch of parameters, so their use is more complex.) Anyway, this does not reduce the complexity of the full algorithm---only that of the hierarchical clustering---because computing dissimilarities has a complexity $O(n^2)$.

As of the fact that the result of hierarchical clustering is a dendrogram, from which clusters need to be defined \emph{ad hoc,} this can be regarded as an advantage rather than a drawback. Dendrograms provide a sort of multi-resolution clustering where one can see not only the main clusters, but also sets of nodes forming clusters within clusters---something that may be very informative for some applications (hence the success of phylogenetic trees in evolutionary biology).

The true disadvantage of hierarchical clustering compared to methods based on modularity or stochastic blocks is not only that choosing the right dissimilarity measure is a problem, but that none of these measures uses a null model to decide whether the dissimilarity found between two entities may be spurious \cite{gower:1986,everitt:2011}---as global methods do. To illustrate the problem, consider the case of words (the entities) in documents (the features). Suppose further that the subject of these documents is `politics'. It is clear that a word like `politician' is likely to appear in many of them; but on the other hand, words like prepositions appear in every single document, so the dissimilarity between, say, the word `of' and the word `politician' will be low regardless of the measure we have chosen. And yet, this low dissimilarity is spurious because there is no meaningful connection between these two words. This is the reason why some datasets require an \emph{ad hoc} pre-processing before one of these algorithms can be applied to the data (for instance, in the processing of texts, it is common to remove words bearing no actual meaning, like articles, prepositions, etc.).

The purpose of this paper is to introduce a random null model for bipartite networks and define a dissimilarity measure between pairs of entities in terms of the statistical significance of the shared and not shared features. This will automatically remove spurious relationships such as the one just described. Combined with, e.g., SLINK, an efficient algorithm for single-linkage clustering \cite{sibson:1973}, it will lead to an $O(n^2)$ algorithm to generate a dendrogram from bipartite datasets. Additionally, as we shall see, the algorithm can be readily extended to the case of one-mode networks.

\section{\label{sec:algorithm}Description of the method and the algorithm}

Consider a set of entities $\mathcal{E}$ and a set of features $\mathcal{F}$ with $|\mathcal{E}|=N_E$ and $|\mathcal{F}|=N_F$ elements respectively. Each entity will have some of these features, so a bipartite network can be defined with nodes $\mathcal{E}\cup\mathcal{F}$ and (bidirectional) links joining entities with their features. The adjacency matrix of such a network has the form
\begin{equation}
\mathbf{A}=
\begin{pmatrix}
0 & \mathbf{B} \\
\mathbf{B}^{\mathsf{T}} & 0
\end{pmatrix},
\end{equation}
where $\mathbf{B}=(b_{ir})$, $i\in\mathcal{E}$, $r\in\mathcal{F}$, is such that $b_{ir}=1$ if entity $i$ has feature $r$ and $b_{ir}=0$ otherwise. Accordingly,
\begin{equation}
n_{ij}=\big(\mathbf{B}\mathbf{B}^{\mathsf{T}}\big)_{ij}, \qquad
m_{rs}=\big(\mathbf{B}^{\mathsf{T}}\mathbf{B}\big)_{rs}, 
\end{equation}
count the number of features that entities $i$ and $j$ have in common, and the number of entities having both features $r$ and $s$, respectively. In particular, $n_i=n_{ii}$ counts the number of features of entity $i$ and $m_{r}=m_{rr}$ counts the number of entities having feature $r$.

With these numbers one can introduce all kinds of dissimilarity measures \cite{everitt:2011,gower:1986} with which to construct a hierarchical clustering and produce a dendrogram for the entities revealing which of them are closer to each other. Instead, we will compute what is the probability that entities $i$ and $j$ have at least $n_{ij}$ features in common if features are assigned randomly
to entities (without any bias).

The probability distribution $p(n_{ij}|n_i,n_j,N_F)$ is obtained as follows. We tag all the $n_i$ elements of set $\mathcal{F}$ that correspond to features of entity $i$, and then draw, randomly and without replacement, $n_j$ features out of the set $\mathcal{F}$. The sought probability is the probability that exactly $n_{ij}$ of these extracted elements are tagged, and it is given by the hypergeometric distribution \cite[\S 2.6]{feller:1968}
\begin{equation}
p(n_{ij}|n_i,n_j,N_F)=\frac{\binom{n_i}{n_{ij}}\binom{N_F-n_i}{n_j-n_{ij}}}{\binom{N_F}{n_j}}.
\end{equation}
What we are interested in is the $p$-value
\begin{equation}
p_{ij}=\sum_{k\geq n_{ij}}p(k|n_i,n_j,N_F).
\end{equation}
This will be our measure of dissimilarity between entities $i$ and $j$.

Interestingly, this is a standard problem is statistics that can be solved by building the contingency Table~\ref{tab:table1} and applying \emph{Fisher's exact test} (FET), for which very efficient algorithms are implemented in widely used programming languages such as Python (\texttt{scipy.stats.fisher\_exact}) and R (\texttt{fisher.test}). The outcome of this test is precisely $p_{ij}$, which allows us to easily build the dissimilarity matrix $\mathbf{D}=(p_{ij})$. Given $\mathbf{D}$, we can apply any agglomerative clustering method using the Lance-Williams algorithm, parametrized as single linkage, for updating the dissimilarity between clusters \cite[ch.~4]{everitt:2011} and get the desired dendrogram representing the clustering structure of the entities.

\begin{table}[]
    \centering
    \renewcommand{\arraystretch}{1.5}
    \begin{tabular}{l|c|c||c|}
        \diagbox[width=18mm,height=8mm]{\hspace*{5mm}$i$}{$j$\hspace*{2mm}\ } & \textbf{features} & \textbf{not features} & \textbf{total} \\
        \hline
        \textbf{features} & $n_{ij}$ & $n_i-n_{ij}$ & $n_i$ \\
        \hline
        \textbf{not features} & $n_j-n_{ij}$ & $N_F+n_{ij}-n_i-n_j$ & $N_F-n_i$ \\
        \hline\hline
        \textbf{total} & $n_j$ & $N_F-n_j$ & $N_F$ \\
        \hline
    \end{tabular}
    \caption{\label{tab:table1}Contingency table to apply Fisher's exact test. Entity $i$ has $n_i$ out a set of $N_F$ features, $n_{ij}$ of which are shared with $j$ and $n_i-n_{ij}$ are not. Thus, $n_j-n_{ij}$ features of $j$ are not shared by $i$, and a total of $N_F+n_{ij}-n_i-n_j$ features are exhibited neither by $i$ nor by $j$.}
\end{table}



Once we obtain a dendrogram, deciding the optimal number of clusters (if any) might not be a trivial matter. Several cluster validity indexes (CVIs) have been proposed as a way of selecting the best number of clusters, usually based on between- and within-cluster distances~\cite{Clustering_validation_1}---in a metric, mathematical sense. Given that our dissimilarity matrix $\mathbf{D}$ is not a matrix of true distances (in the strict sense), we propose a different approach. In particular, we will choose the partition that maximizes the \emph{susceptibility} of the system as used in percolation theory. This susceptibility is formally defined as $\chi=\sum{n_{s}s^{2}/N}$, where $n_{s}$ is the number of clusters of size of size $s$, and the sum is taken over all but the largest cluster (see~\cite{PhysRevE.90.052810} for more details).

We prefer susceptibility---a measure taken from statistical mechanics---rather than any of the more standard CVIs developed in computer science because, as the percolation example shows, susceptibility is very sensitive to the breaking of a big cluster into smaller ones---precisely the problem we face here. Although several other measures (like silhouette coefficient, Calinski–Harabasz index, or Davies–Bouldin index) could be used, we convinced ourselves that susceptibility is indeed more informative than several  by comparing the outcomes in a few examples.

As a technical comment, notice that the maximum possible value of $\chi$ is achieved when the network breaks into two equally sized clusters, yielding $\chi_{max}=N/4$. To make this measure independent of the network size---hence comparable---we use $\chi=4\sum{n_{s}s^{2}}$ as our normalized susceptibility.

For the sake of consistency, we will use $\chi$ to select the optimal number of clusters in all the case studies reported in this manuscript. This notwithstanding, we would like to recall that the multi-resolution nature of hierarchical clustering still provides useful information, and that the actual \emph{best} partition of particular data is a problem-dependent question.

\section{Data Analysis}

We test the performance of our algorithm (henceforth referred to as \emph{clusterBip}) with different datasets. Firstly, we generate synthetic networks with a well-defined community structure and challenge the algorithm to uncover it. Secondly, we use real-world data from congressional voting records (U.S.A) and try to classify the Congressmen in their corresponding political parties---which act as background truth for the underlying community structure. Lastly, we use data from a massive survey carried out in France in 2003 to analyze how some leisure activities are more related than others. 

\subsection{Computer-Generated Networks}
 
We begin by analyzing bipartite networks created synthetically with a community structure established \emph{ex ante}. All of these networks will consist of 100 entities and 400 features, connected following different heuristics so that we have a reliable background truth with which to compare the results provided by the algorithm.

\begin{figure*}
\centering
\includegraphics[width=\textwidth]{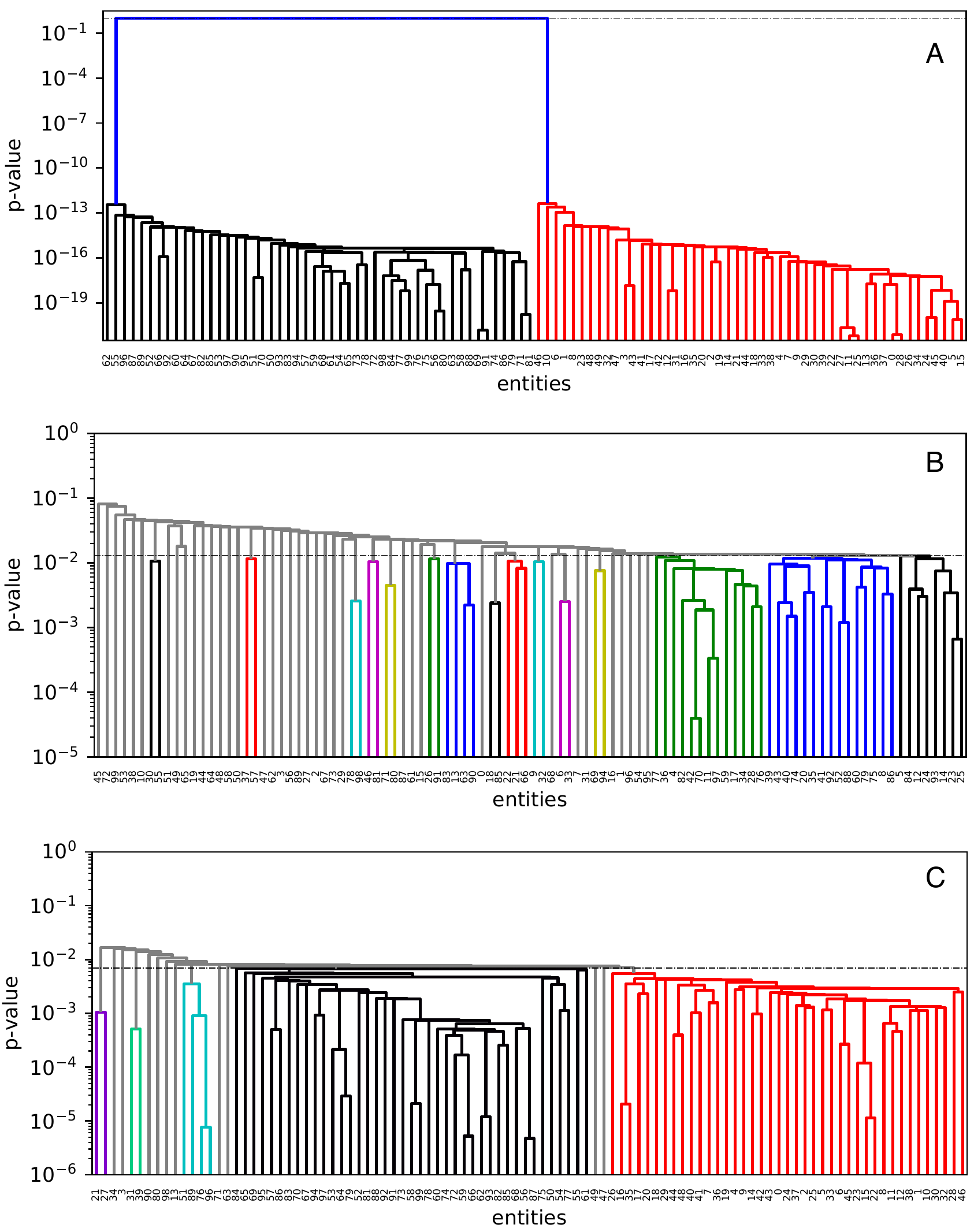}
\caption{\label{fig:computerGeneratedNets}
Analysis of synthetic bipartite networks. (A) dendrogram of a two-cluster bipartite network. (B) dendrogram of a random network. (C) dendrogram for an intermediate case with $p_{\rm add}=0.28$. In (A), (B), and (C) the dashed lines mark the point with highest susceptibility---that where the `optimal' partition should be found.}
\end{figure*}

The most extreme case of a bipartite network with community structure is a network created as the union of two separate (bipartite) ones. To build such a network, we first select 50 entities and 200 features and create a link between any two of them with probability $1/2$. Then, we take the remaining nodes (50 entities and 200 features) and proceed similarly. When these two networks are put together, the result is, by construction, a two-cluster bipartite network. In Fig.~\ref{fig:computerGeneratedNets}A we can see how clusterBip captures this situation seamlessly. The entities are grouped into two different clusters (red and black), which exactly correspond to the original building blocks of the network. Furthermore, the (normalized) susceptibility peaks at $p$-value $p=1$, where the two clusters split, and achieves its maximum possible value, $\chi=1$ (see Fig.~S1 in the SM). Notice that some weak structure can be observed within the two main clusters. It is just caused by random fluctuations~\cite{PhysRevE.70.025101}, and the low values of $\chi$ (SM, Fig.~S1) confirm this fact.

\begin{figure}
\centering
\includegraphics[width=\columnwidth]{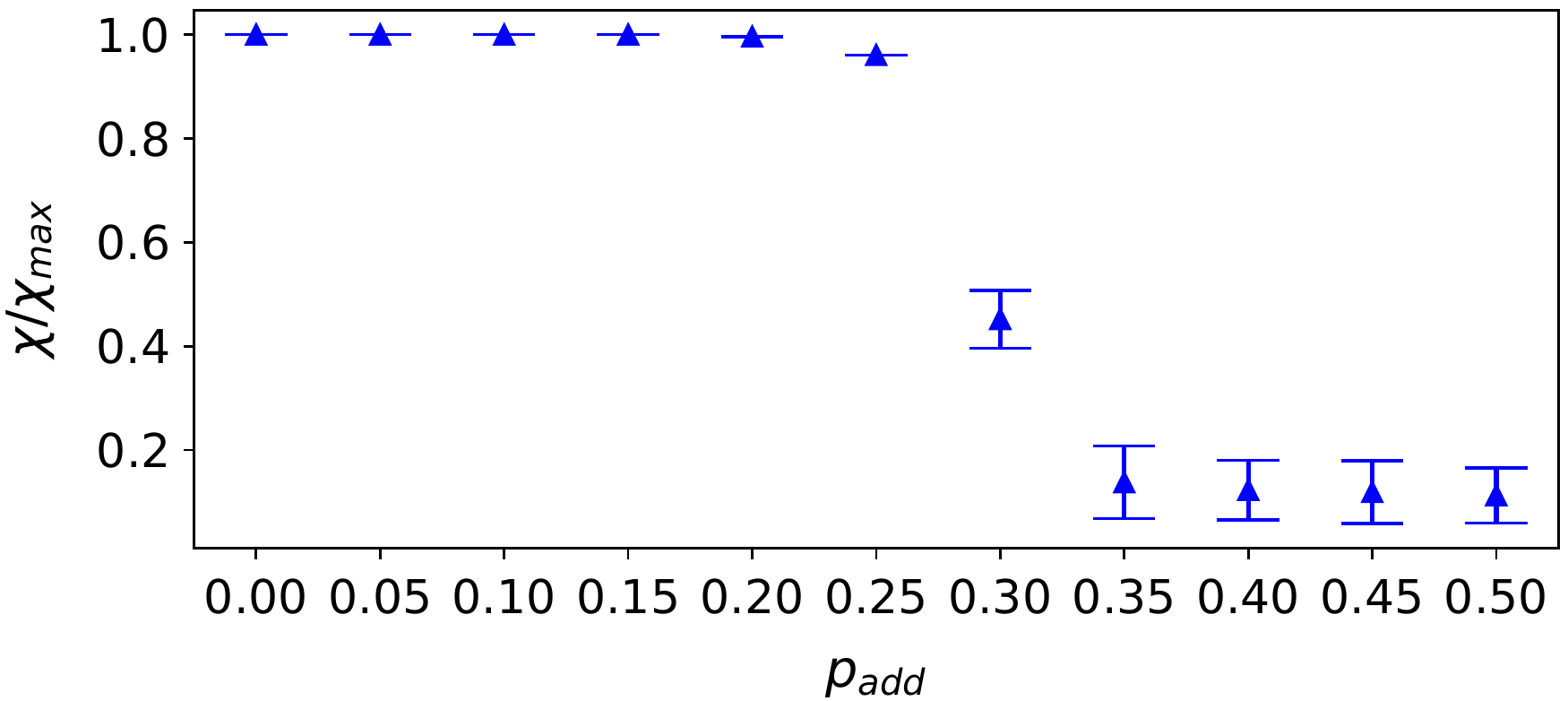}
\caption{\label{fig:computerGeneratedNets2}
Analysis of synthetic bipartite networks. Maximum value of the normalized susceptibility $\chi/\chi_{\rm max}$ (which suggests the point of optimal partition) as a function of $p_{\rm add}$, the probability that a link connects with a node at the `wrong' module. For each value of $p_{\rm add}$ we have generated $100$ realizations of the networks. The values of the susceptibility are the averages over these realizations and the error bars indicate the corresponding standard deviations.}
\end{figure}

We now apply the algorithm to a purely random network. We generate it from the previous two-cluster network by adding links with probability $1/2$ between the entities of each block and the features of the other. The identity of the two blocks has thus disappeared, and any pattern observed should be spurious. The resulting dendrogram is depicted in Fig.~\ref{fig:computerGeneratedNets}B. Just a glimpse to this figure reveals that there is no clear structure---something that the low values of the susceptibility ($\chi=0.134$) at the threshold point ($p=0.013$) confirm (SM, Fig.~S2). The high $p$-value at the threshold also confirms that the clustering has low statistical significance.

These results show that clusterBip performs well for the two extreme cases that we have devised. But we can also test it for intermediate cases. We generate these intermediate benchmarks by connecting nodes of opposite modules with varying probability $0<p_{\rm add}<1/2$ ($p_{\rm add}=0$ would correspond to the two-block network and $p_{\rm add}=1/2$ to the random network). Figure~\ref{fig:computerGeneratedNets}C shows what the clusters formed looks like for an intermediate case ($p_{\rm add}=0.28$). The identity of a few nodes can no longer be recovered, but the two cluster are still clearly identifiable. Figure~\ref{fig:computerGeneratedNets2} shows the largest value of the susceptibility for several values of $p_{\rm add}$, which identifies the point of the `optimal' partition for each network. The susceptibility remains close to $1$ up to $p_{\rm add}\approx 0.25$, indicating that the two modules are clearly identified even if one fourth of the links connect to the `wrong' module. Then the susceptibility undergoes a sharp decay and beyond $p_{\rm add}\approx 0.35$ it practically vanishes---as for the random network. Fig.~\ref{fig:computerGeneratedNets}C illustrates a case within this region of sharp decay of the susceptibility. What we see in this figure is representative of what happens, that is, the identity of the two clusters degrades as $p_{\rm add}$ increases.

We have also tested the robustness of these results for networks with heterogeneous degree distribution. To this purpose we have generated the two blocks as bipartite networks with power-law degree distribution using the preferential attachment algorithm (function \verb+preferential_attachment_graph+ from the NetworkX Python package \cite{networkx}). The results are similar---clusterBip is able to capture the network structure regardless of the degree distribution of the network (see Fig.~S22 in the SM).

\subsection{Congressional Voting Records}
\label{sec:voting}

The U.S. Congressional voting records dataset \cite{Voteview} gathers all roll call votes made by the United States Congress during the years 1789-2017. Each Congressman is represented by a set of features describing how he voted on every bill for a Chamber (Senate or House), for a particular Congress (period of two years). There are nine different ways of voting---the so-called `cast codes' (see ref.~\cite{Voteview} for more details). Prior to analyzing the data, we process them as in~\cite{Porter7057}, that is, we group cast codes 1, 2, and 3 (ways of voting `yea'), cast codes 4, 5, and 6 (ways of voting `nay'), and cast codes 0, 7, 8, and 9 (not voting). As a result, we build bipartite networks which consist of Congressmen (entities) linked to their particular vote (`yea', `nay', or `not voting') on a particular bill (features).

It is well-known that Congressmen usually vote according to their political party commandments---more so in the recent period \cite{andris:rise}. Hence, clusterBip should be able to determine the political party of the different Congressmen based on how they voted the different bills. To test this hypothesis, we analyze the Congressional voting records of both the House and Senate for Congresses 114th (years 2015-2016) and 36th (years 1959-1960).

\begin{figure*}
\centering
\includegraphics[width=\textwidth]{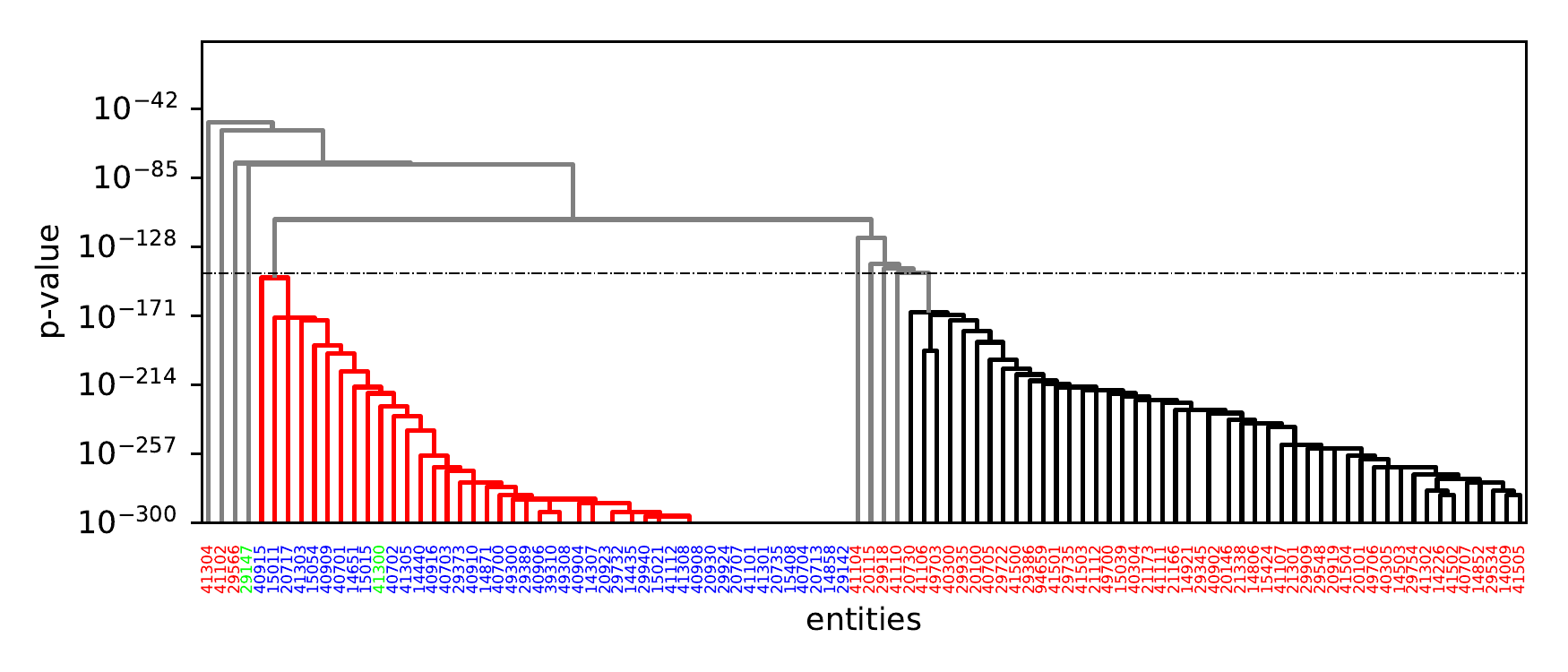}
\caption{\label{fig:S114_dendro}
Dendrogram of the clustering of roll cast votes of Congressmen for Senate 114. Colors red and black in the dendrogram identify clusters of Congressmen; gray color is used for Congressmen not assigned to any cluster. Labels in horizontal axis identify Congressmen and their color identifies the political party they belong to. (The dendrogram is cut at $p$-values around $10^{-300}$ because below this value the computation yielded underflows.)}
\end{figure*}

The results for both Congresses confirm our hypothesis. The algorithm detects two main clusters, and these clusters correspond to the different political parties (Democrats and Republicans). In Fig.~\ref{fig:S114_dendro} we present results for Senate 114, in which Congressmen are more polarized (exhibiting higher values of susceptibility, Fig.~S12 of the SM) and the number of Congressmen is smaller than those in House's datasets---hence making it more suitable for visual interpretation of the dendrogram. We predict Congressmen groups (membership to party) with a 92\% of accuracy for Senate 114, 98.41\% for House 114 (SM, Figs.~S6 and S7), 88.57\% for Senate 36 (SM, Figs.~S8 and S9), and 78.37\% for House 36 (SM, Figs.~S10 and S11). As a matter of fact, this lower accuracy in the results for the 36th Congress is not attributable to a lower performance of the algorithm, but to the higher polarization trend that Congressmen have undergone over the years \cite{andris:rise}. The effect is also observed in other clustering techniques (see Sec.~\ref{sec:comparison}; see also Figs.~S13-S16 in the SM for a multiple correspondence analysis \cite{abdi2007multiple} of the same data).

\subsection{``Life story'' dataset}
\label{sec:surveys}

\begin{figure*}
\centering
\includegraphics[width=\textwidth]{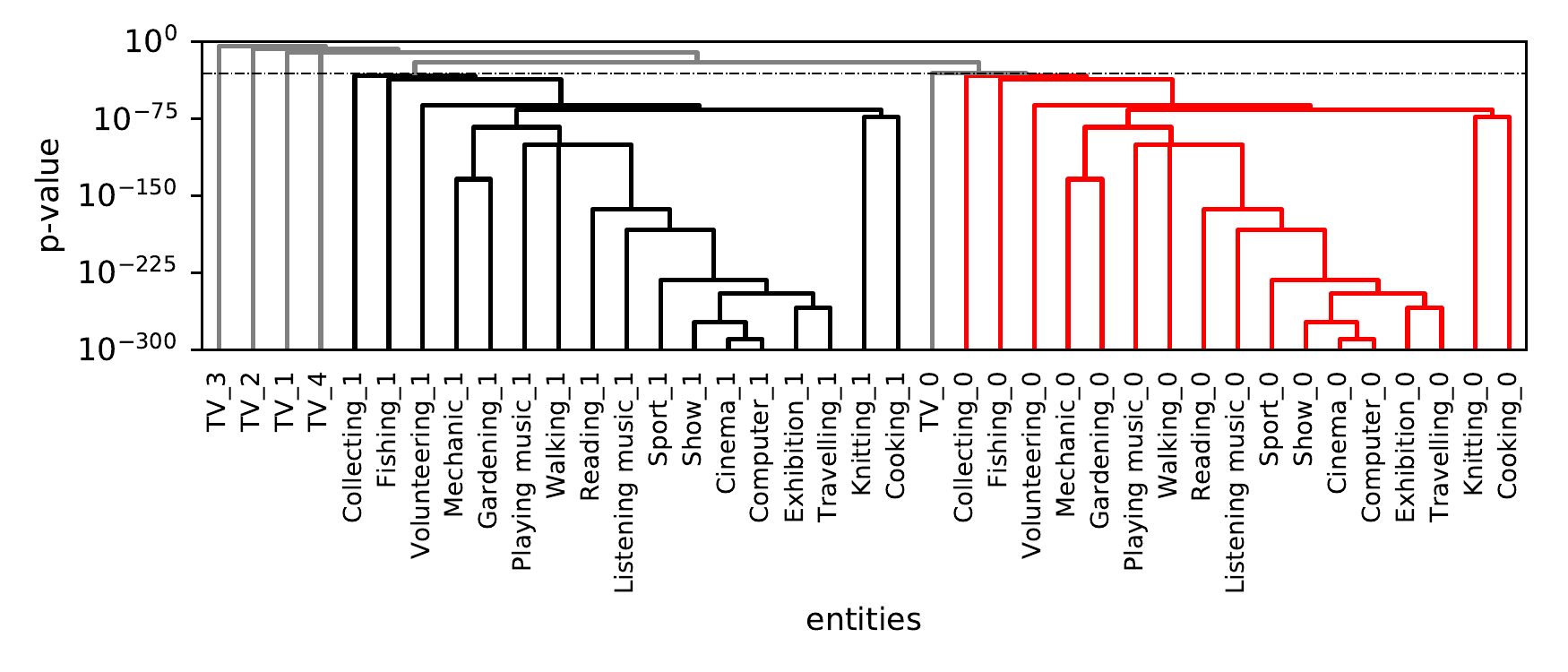}
\caption{\label{fig:Hobbies_dendrogram}
Dendrogram of the clustering of leisure activities in the ``life story'' dataset. The dashed line marks the $p$-value with the largest susceptibility.}
\end{figure*}

We illustrate another application of the method by analyzing data from the 2003 INSEE (Institut National de la Statistique et des \'Etudes \'Economiques) survey on identity construction, the so-called ``life story'' survey~\cite{LifeStory}. The study was conducted in the metropolitan areas of France and recorded answers are related to family and professional situation, geographic and social origins, ethical commitments, cultural practices, and state of health. In particular, the dataset we analyze comprises the answers of $8403$ people ($55\%$ female) to the question \emph{`Which of the following leisure activities do you practice regularly?'}, and the answer choices were: \emph{Reading}, \emph{Listening to music}, \emph{Cinema}, \emph{Shows}, \emph{Exhibitions}, \emph{Computer}, \emph{Sport}, \emph{Walking}, \emph{Travel}, \emph{Playing a musical instrument}, \emph{Collecting}, \emph{Voluntary work}, \emph{Home improvement}, \emph{Gardening}, \emph{Knitting}, \emph{Cooking}, \emph{Fishing}, \emph{Number of hours of TV per day on average} (0-4). In addition to this information, the data includes four supplementary variables: \emph{sex}, \emph{age}, \emph{profession}, and \emph{marital status} (see Figs.~S17 and S18 of the SM for a brief summary of descriptive statistics). The dataset is available within the R package \emph{FactoMineR}~\cite{factominer}.

With these data, we build a bipartite network of people (entities) and leisure activities (features) and apply the algorithm to find groups of activities whose co-appearance in the individuals' answers is statistically significant. Each of the activities is represented by two nodes in the set of features, one corresponding to practicing it, and the other to not doing so. That way, we can get clusters that include doing some activities and not doing some other ones. The dendrogram represented in Fig.~\ref{fig:Hobbies_dendrogram} exhibits two main clusters: the black one groups all the nodes corresponding to actively performing activities, whereas the red one collects all the nodes corresponding to not performing them. On the other hand, watching TV does not seem to be particularly related to any of these clusters. Therefore, the main contrast between the different leisure activities is, precisely, whether they are actively performed or not.

\begin{figure*}
\centering
\includegraphics[width=\textwidth]{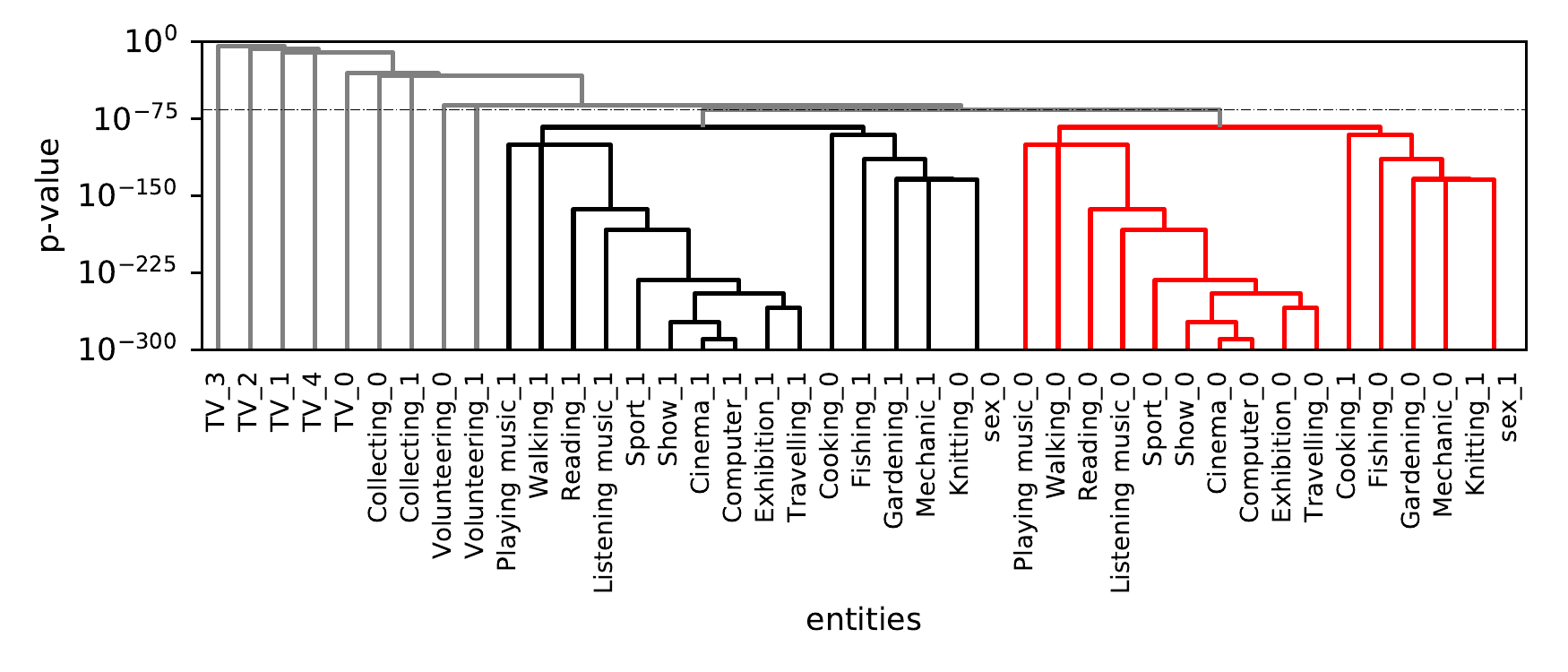}
\caption{\label{fig:hobbies_dendro_withSex}
Same as Fig.~\ref{fig:Hobbies_dendrogram}, but including the supplementary variable `sex'.}
\end{figure*}

These two clusters are identified by the peak of the susceptibility. However, each of them is further structured in interesting subgroups. For instance, watching movies at the cinema and spending time with the computer appear as two very closely related activities. Moving one step above in the hierarchical dendrogram we see that these two activities are also commonly performed by people who enjoy attending to shows and, moving even one step further, by people who enjoy exhibitions and traveling.

More relevant information is gained if we include the variables \emph{male} (\emph{sex} $=0$) and \emph{female} (\emph{sex} $=1$) as features. Figure~\ref{fig:hobbies_dendro_withSex} shows the resulting dendrogram. We can observe that the activity more closely related to sex is knitting, which seems to be predominately practiced by females. Not only that: a bit below the first splitting into two main clusters (at the peak of the susceptibility) there is a secondary splitting that associates fishing, gardening, mechanics, not cooking, and not knitting with males (and the opposite with females), whereas activities such as walking, reading, listening to music, practicing sports, going to shows, exhibitions or the cinema, using computers, and traveling, form a cluster weakly related to sex. Lastly, the analysis also reveals that activities such as collecting or volunteering are not preferred by any particular sex.


For the ``life story'' data set there is no absolute background truth that we can use to validate our results. However, we can compare them with what is obtained applying one of the most commonly used techniques for analyzing the structure of categorical data: multiple correspondence analysis (MCA), an adaptation of standard correspondence analysis to categorical data~\cite{abdi2007multiple}. Like principal component analysis, MCA represents the data as points in a low-dimensional Euclidean space that retains the maximum variance of the data. A Chi-square test is used to examine whether rows and columns of a contingency table are statistically significantly associated, and component analysis decomposes the chi-squared statistic associated with this table into orthogonal factors (dimensions). Eventually, MCA can be used to find groups of categories (features) or individuals (entities) that are similar.

\begin{figure}
\centering
\includegraphics[width=\columnwidth]{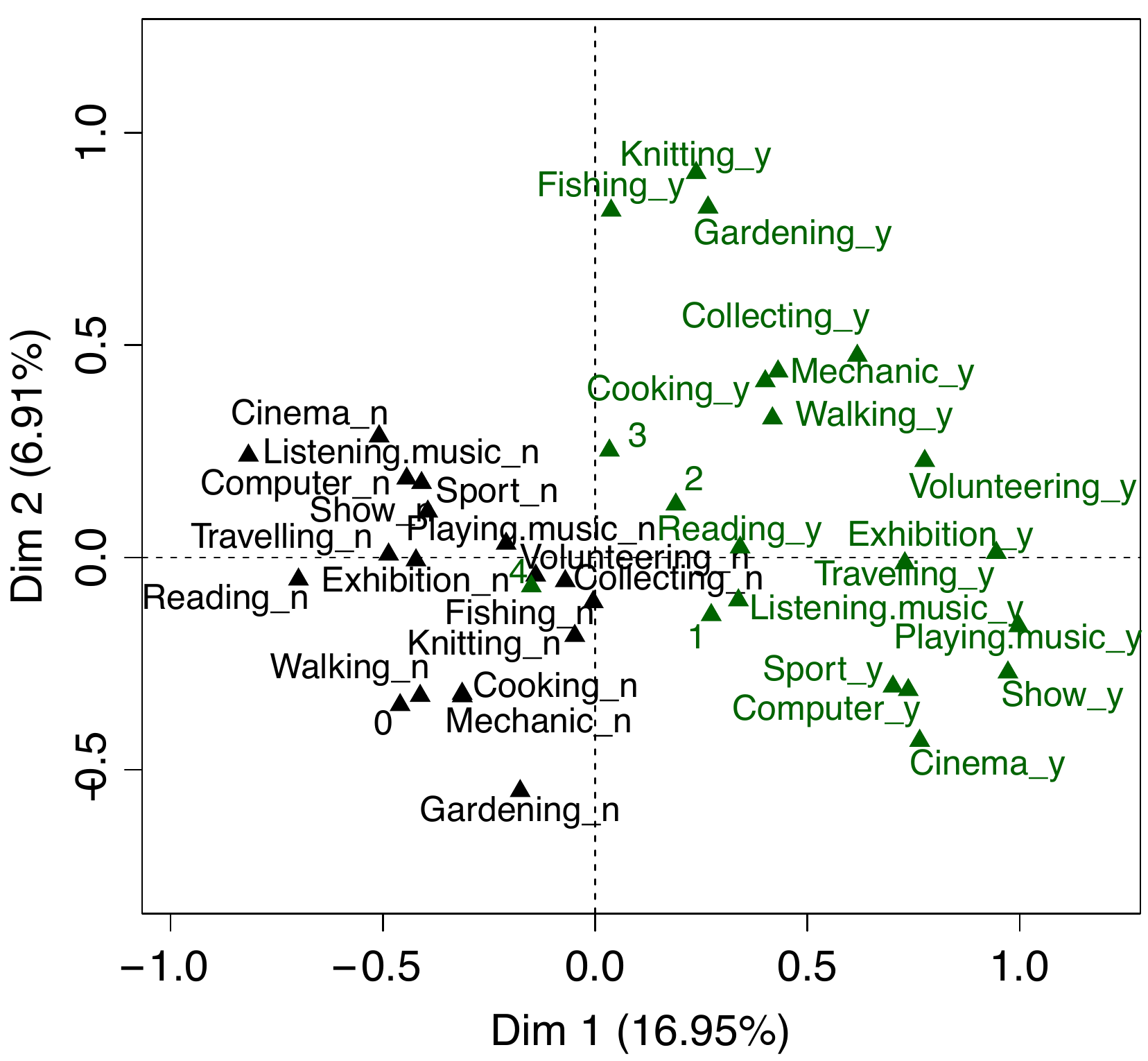}
\caption{\label{fig:MCA_hobbies}Factors map of the ``life story'' data set. In green, the activity is performed; in black, the activity is not performed.}
\end{figure}

We performed MCA on the ``life story'' data set using the R package \emph{FactoMineR}~\cite{factominer}. In the so-called factors map (see Fig.~\ref{fig:MCA_hobbies}), the distance between two features is a measure of their dissimilarity. Each feature is represented at the barycenter of the individuals in it. Features with a similar profile are grouped together, and negatively correlated features appear on opposite sides of the plot origin (opposed quadrants). The distance between feature points and the origin measures the quality of the feature points on the factor map. Feature points that are away from the origin are well represented on the factor map. In our example, this representation only captures a 24\% of the variance of the data set.

Grouping features in a MCA factor map is rather subjective. In Fig.~\ref{fig:MCA_hobbies} we have represented our data points using two different colors: green for actively performed activities, and black for non-performed ones. The points seem to be cluttered in two groups, separated by an imaginary diagonal going from the second to the fourth quadrant. This separation is congruent with the two main clusters detected by our algorithm (see Fig.~\ref{fig:Hobbies_dendrogram}), signaling that this division is indeed present in the data. However, looking for further structure using the factor map is rather complicated. Let us recall that a factor map is a two-dimensional representation of the data that only captures part of its variance ($24\%$ in this case). Hence, the distance between any two points in the map can not be trusted as a measure of their true dissimilarity. 


Unlike MCA, clusterBip does not lose information by projecting into a lower dimensional space, and takes into account the fact that co-occurrences between features can be due to chance (unlike other hierarchical methods \cite{gower:1986,everitt:2011}). Hence, even though both techniques are coincident in their coarse grain classification of the data, the method we introduce here is able to unveil finer substructure that remains hidden in the MCA map, and can provide deeper insights when analyzing survey data---or any other data with bipartite structure.

\section{Application to one-mode networks}
\label{sec:football}
\begin{figure*}
\centering
\includegraphics[width=\textwidth]{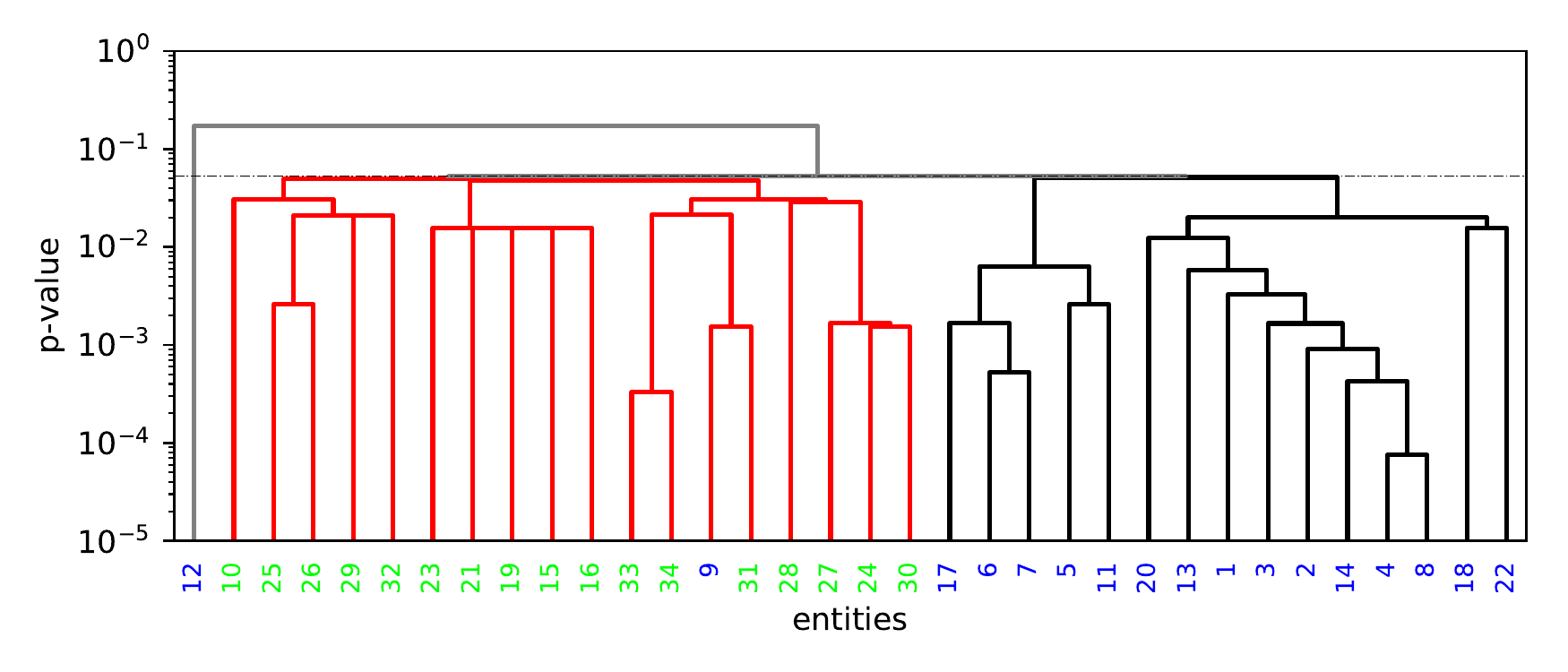}
\caption{\label{fig:karate}
Dendrogram of Zachary's Karate Club network. The dashed line corresponds to the point with the largest susceptibility.}
\end{figure*}

\begin{figure*}
\centering
\includegraphics[width=\textwidth]{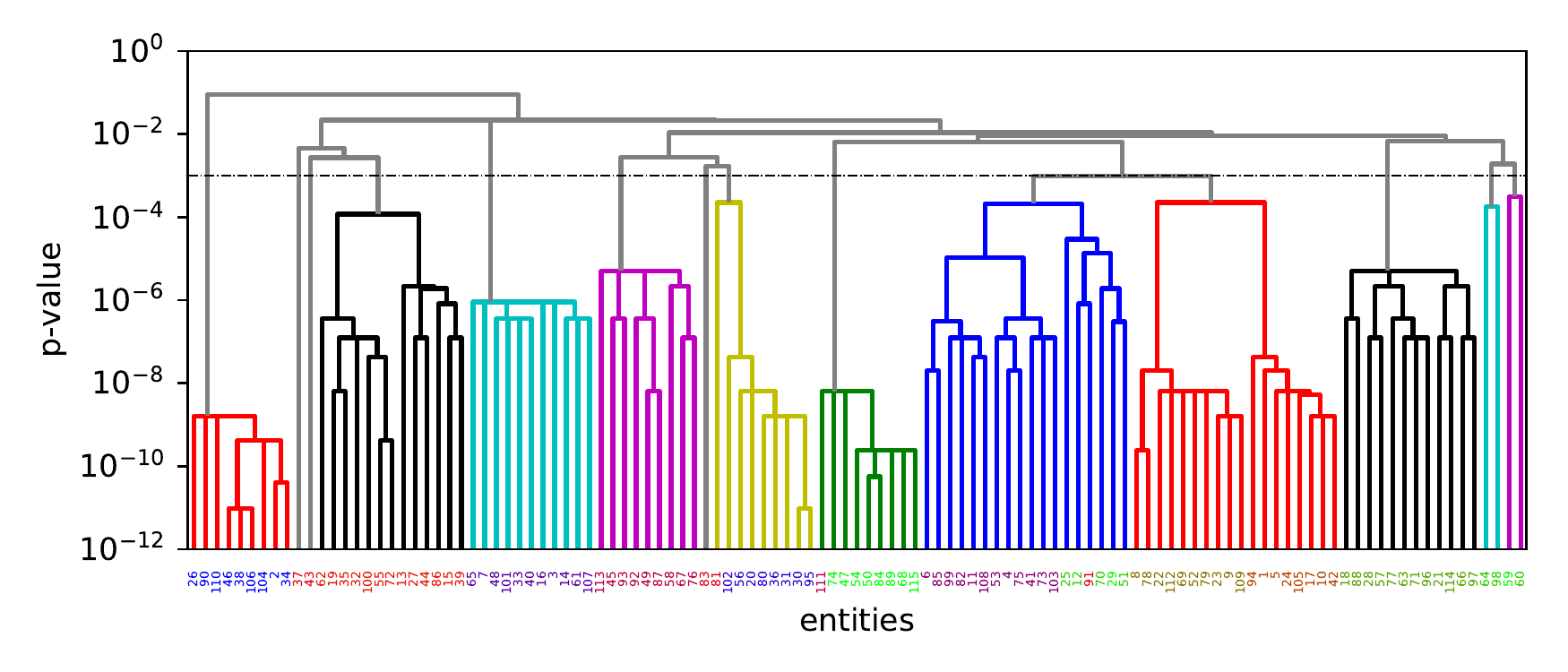}
\caption{\label{fig:football} Dendrogram of the American Football dataset. The dashed line corresponds to the point with the largest susceptibility.}
\end{figure*}

As a final application of clusterBip, we will illustrate how it can be applied to ordinary, one-mode networks. As we discussed in the introduction, when trying to analyze bipartite networks, a typical procedure is to project them onto one-mode ones, so that one can take advantage of the many techniques developed for this type of graphs~\cite{zhou:2007} (alternatively, these techniques can be sometimes extended to be directly applied to two-mode networks~\cite{everett2013dual}). What is rarely done is the reverse procedure, that is, to create a bipartite network out of a one-mode one in order to benefit from techniques initially tailored for the former. We propose here a simple way to do this. What we gain from doing so is that we can exploit the faster performance of our algorithm to quickly produce a multi-resolution clustering of an ordinary network.

Consider the network $\mathcal{G}$ with a set of nodes $\mathcal{N}$ and adjacency matrix $\mathbf{C}$---which we consider symmetric with zero diagonal (no self-loops) for simplicity. Now, we identify $\mathcal{N}$ with the set of entities, $\mathcal{E}$, create a replica of the same set, and identify it with the set of features, $\mathcal{F}$. Links joining nodes of $\mathcal{N}$ now join nodes of $\mathcal{E}$ with its neighboring nodes in the replica $\mathcal{F}$, thus transforming the original network $\mathcal{G}$ into a bipartite network. Furthermore, in order to eliminate the possibility for this bipartite network to be disconnected, we need to link each node of $\mathcal{E}$ with its own replica in $\mathcal{F}$. Therefore, the bipartite network will be described by the adjacency matrix
\begin{equation}
\mathbf{A}=
\begin{pmatrix}
0 & \mathbf{I+C} \\
\mathbf{I+C} & 0
\end{pmatrix}.
\end{equation}

In order to reveal the statistical meaning of the adjacency matrix so obtained notice that
\begin{equation}
n_{ij}=\big((\mathbf{I}+\mathbf{C})^{2}\big)_{ij}=\delta_{ij}+2\mathbf{C}_{ij}+\big(\mathbf{C}^{2}\big)_{ij}.
\end{equation}
Now if $\mathcal{V}_i$ denotes the set of nodes at distance at most one from node $i$ (the \emph{neighborhood} of node $i$, including itself), then $n_{ij}=|\mathcal{V}_i\cap\mathcal{V}_j|$, in other words, $n_{ii}$ is the size of $\mathcal{V}_i$ and $n_{ij}$, for $i\ne j$, counts the number of common nodes in the neighborhoods of nodes $i$ and $j$. Accordingly, the dissimilarity of nodes $i$ and $j$ is a measure of the statistical significance of the overlap of their neighborhoods with respect to the null model provided by the configuration model---with neither loops nor multiple links.



To illustrate this application of the algorithm, we turn to two well-known benchmarks, one-mode networks with community structure: the Zachary’s karate club study \cite{zachary:1977,girvan:2002}, and the College Football dataset \cite{girvan:2002}. In the first of them, Zachary monitored the relationships of $34$ individuals attending a Karate club that eventually split into two different ones. Very often in the literature, the performance of community detection algorithms is assessed by how well they predict this partition \cite{fortunato:2016}. Our algorithm accomplishes this task almost perfectly, classifying incorrectly only two nodes. One of them is node 9, which Zachary himself misclassified in his study \cite{zachary:1977}. Furthermore, in Fig.~\ref{fig:karate} we can detect at least one clear subgroup composed by nodes 17, 6, 7, 5, and 11 (all in black), which is also captured by other classic algorithms for one-mode networks \cite{girvan:2002}.



The College Football network is formed by a set of 115 College Football teams (nodes) which are connected to each other if they were confronted during the regular-season of the Division I in 2000 (U.S.A.)~\cite{girvan:2002}. In reality, the different teams are divided into what is known as conferences, each containing between 8 and 12 teams. Intra-conference games are more common than inter-conference ones, so teams belonging to the same conference are highly interconnected in the network, and a community detection algorithm should be able to account for this. In Fig.~\ref{fig:football} we can appreciate how our algorithm does so. The labels at the horizontal axis represent the different clubs, colored according to the conferences they belong to. As we can see, most of them are grouped together under the same branch in the dendrogram, which is able to uncover the structure of the conferences. Let us note that the branches are colored according to the partition that maximizes the susceptibility, which should be taken as a guidance, but that, as in this case, it might not correspond to the (actual) best partition (see also the discussion in section \ref{sec:algorithm}).

\section{Comparison with state-of-the-art algorithms}
\label{sec:comparison}

As we have explained before, a dendrogram provides much more information at several scales that a simple partition into clusters. Despite this, it can be informative to have a comparison with some state-of-the-art, clustering methods. Since clusterBip can be applied both to bipartite and one-mode networks, the class of algorithms we can compare it with is different in each case. Thus we perform two separated analyses.

As bipartite networks, we have analyzed the Southern Women (S.W.) dataset \cite{davis2009deep} (which records the attendance at 14 social events by 18 Southern women) both ways: clustering women and clustering events; the two cases of the Congressional voting records dataset---a heavy politically polarized house (H114 for House 114) and a more diverse house (H036 for House 36)---of Sec.~\ref{sec:voting}; and the ``Life story'' dataset of Sec.~\ref{sec:surveys} (Hobbies and Hobbies*, which includes gender). Table~\ref{table:bip1} shows the number of clusters found by Brim \cite{barber:2007}, Rnetcarto \cite{guimera:2007}, Infomap \cite{Rosvall1118}, and clusterBip for each of these dataset. It is striking the poor performance of Infomap when applied to projected bipartite networks, probably because projection looses too much relevant information in the process \cite{Alzahrani2016}. Leaving Infomap aside, overall the number of clusters found by clusterBip is similar to those found by the other two methods. The Southern Women dataset is an exception when applied to cluster women, because clusterBip detects 5 clusters where Brim and Rnetcarto detect 3 and 2, respectively. However, the disagreement between these two methods and the fact that clusterBip detects women that cannot be clustered reveals that any clustering of women in this dataset is doubtful and may respond to some cues and not others. A dendrogram in this case provides more insights into the classification (more details can be found in the captions of Figs.~S24 and S25 of the SM).

\begin{table}[htb]
\centering
\begin{tabular}{lcccc}
\hline
 & Brim & Rnetcarto & Infomap & clusterBip \\
\hline
S.W. (women)  & 3  & 2  & 2  & (5, 3)  \\
S.W. (events) & 3  & 2  & 1  & (2, 1)  \\
H114          & 2  & 3  & 1  & (2, 8) \\
H036          & 3  & 3  & 1  & (3, 6)  \\
Hobbies       & 1  & 2  & 1  & (2, 5)  \\
Hobbies*      & 4  & 2  & 1  & (2, 5) \\
\hline
\end{tabular}
\caption{Comparison of the number of clusters found by clusterBip with those found by two state-of the-art community detection algorithms for bipartite networks (Brim \cite{barber:2007} and Rnetcarto \cite{guimera:2007}), as well as Infomap \cite{Rosvall1118}, an algorithm for one-mode networks that can be applied to the weighted projected network of the original bipartite one. Datasets considered: the Southern Women (S.W.) dataset \cite{davis2009deep} (both ways, women and events), the two Congressional voting records dataset of Sec.~\ref{sec:voting}: House 114 (H114) and House 36 (H036), and the ``Life story'' dataset of Sec.~\ref{sec:surveys} (Hobbies and Hobbies*, which includes gender).}
\label{table:bip1}
\end{table}

\begin{table}[htb]
\centering
\begin{tabular}{lcccccccc}
\hline\small
 & \multicolumn{2}{c}{Brim} & \multicolumn{2}{c}{Rnetcarto} & 
 \multicolumn{2}{c}{clusterBip} \\
\hline
 & NMI & ARI & NMI & ARI & NMI & ARI \\
H114 & 0.98 & 0.99 & 0.99 & 0.99 &
0.93 & 0.97 \\
H036 & 0.65 & 0.69 & 0.68 & 0.67 &
0.68 & 0.73 \\
\hline
\end{tabular}
\caption{Normalized Mutual Information (NMI) and Adjusted Rand-Index (ARI) of the clusters obtained for the Congressional Voting Records dataset, using the political party of the congressmen as ground truth. We compare the results of two common community-detection methods for bipartite networks (Brim \cite{barber:2007} and Rnetcarto \cite{guimera:2007}), as well as of clusterBip. Infomap \cite{Rosvall1118} has been excluded from the comparison because it is unable to detect clusters in this dataset (see Table~\ref{table:bip1}).}
\label{table:bip2}
\end{table}

\begin{table*}[ht]
\centering
\begin{tabular}{lccccccccccccccccc}
\hline\small
 & \multicolumn{2}{c}{ground truth} & \multicolumn{3}{c}{clusterBip} & \multicolumn{2}{c}{Louvain} & \multicolumn{2}{c}{Fastgreedy} & \multicolumn{2}{c}{Infomap} & \multicolumn{2}{c}{Eigenvector} & \multicolumn{2}{c}{LP} & \multicolumn{2}{c}{Pereda et. al} \\
\cline{2-18} 
Networks & C & Q & NMI & Q & $\chi$ & NMI & Q & NMI & Q & NMI & Q & NMI & Q & NMI & Q & NMI & Q \\
\hline
Karate &  2 & 0.37 & 0.78 & 0.36 & 0.78 & 0.59 & 0.42 & 0.69 & 0.38 & 0.70 & 0.40 & 0.68 & 0.39 & 0.70 & 0.40 & 1.00 & 0.37 \\
Polbooks &  2 & 0.41 & 0.56 & 0.47 & 0.62 & 0.51 & 0.52 & 0.53 & 0.50 & 0.49 & 0.52 & 0.52 & 0.47 & 0.57 & 0.50 & 0.89 & 0.38 \\
Football & 12 & 0.55 & 0.89 & 0.55 & 0.33 & 0.88 & 0.60 & 0.70 & 0.55 & 0.92 & 0.60 & 0.70 & 0.49 & 0.92 & 0.60 & 0.91 & 0.66 \\
Dolphins &  2 & 0.38 & 0.31 & 0.38 & 0.21 & 0.48 & 0.52 & 0.61 & 0.50 & 0.50 & 0.52 & 0.54 & 0.49 & 0.69 & 0.50 & 0.60 & 0.44 \\
Polblogs &  2 & 0.41 & 0.18 & 0.20 & 0.04 & 0.63 & 0.43 & 0.65 & 0.43 & 0.48 & 0.42 & 0.69 & 0.42 & 0.69 & 0.43 & 0.71 & 0.52 \\
\hline
\end{tabular}
\caption{Comparison of modularity (Q) and Normalized Mutual Information (NMI) of the clusters detected by clusterBip as well as by six different standard community-detection techniques, when applied to five common one-mode-network benchmarks: Zachary's karate club \cite{zachary:1977}, Polbooks \cite{polbooks}, Football \cite{girvan:2002}, Dolphins \cite{dolphins}, and Polblogs \cite{polblogs}. The second and third columns show the number of clusters (C) and the modularity (Q) of the ground truth of the networks. In the case of clusterBip, we also include the value of the susceptibility $\chi$ at the cut point. Data for the twelve right columns are taken from Ref.~\onlinecite{PEREDA2019320}.}
\label{table:one-mode}
\end{table*}

One interesting feature of clusterBip that provides an advantage over its alternatives is that fact that some nodes remain unclassified. This is the case, for example, of the Congressional voting records datataset: we found 8 congressmen in H114 and 6 in H036 whose voting behavior does not align well with any of the two political parties, which suggests that these congressmen act with some degree of independence. If a party were looking for favorable votes from the opposite party, these are the candidates they should focus on. 

Another example can be found in the clustering of events in the Southern Women dataset (see Figs.~S26 and S27 of the SM). Although clusterBip detects two clusters, just as Rnetcarto, it also singles out one particular event as not belonging to any cluster. It turns out that this is a popular event which women of both cluster attended at, so any coincidence with other events is therefore not significant. It is precisely eliminating spurious relations what clusterBip is particularly good at.

In table~\ref{table:bip2} we analyze again the H114 and H036 datasets, assuming as ground truth the membership to a party of the congressmen. We evaluate the clustering results by computing the Normalized Mutual Information score (NMI)~\cite{NMI} and the Adjust Rand Index (ARI)~\cite{ARI}. In spite that clusterBip cannot score 1 in any of these indexes because of the existence of unclassified nodes, its performance is comparable to that of Brim or Rnetcarto (Infomap is excluded from this comparison because it fails to find any cluster in these datasets). It follows that there is a strong coincidence in the clusters found by all three methods. However, clusterBip is providing more insightful information: it is questioning the very validity of the ground truth by revealing congressmen that do not fit well in either cluster.

As for one-mode networks, we have compared the performance of clusterBip with that of six standard community-detection algorithms, namely Louvain \cite{Blondel2008}, Fastgreedy \cite{fastgreedy}, Infomap \cite{Rosvall1118}, Eigenvector \cite{Eigenvector}, LP \cite{LabelProp}, and Pereda et al.'s~\cite{PEREDA2019320}, when applied to five standard such networks whose ground truth is known: Zachary's karate club \cite{zachary:1977}, \emph{Polbooks,} a network of books about politics which are frequently bought together \cite{polbooks}, the College Football network of Sec.~\ref{sec:football}, a network of associations of bottlenose dolphins \cite{dolphins}, and \emph{Polblogs,} the political blogosphere \cite{polblogs}. For each algorithm and dataset we have computed the modularity (Q) of the clustering and of the ground truth, as well as the NMI score. The results are collected in Table~\ref{table:one-mode}. The performance of clusterBip is comparable to the best results of the other methods, except for the networks Dolphins and Polblogs. In both networks, the small value of the susceptibility $\chi$ questions the existence of a clear community structure. For the case of Polblogs, clusterBip detects two clusters of 153 and 127 nodes, as well as 992 unclassifiable nodes.

All in all, we see that clusterBip performs as well as other state-of-the-art methods, but is able to provide both, further structure within clusters, and sets of nodes that do not belong in any of the detected clusters. Often this information is as important as that of the clustering itself.

\section{Discussion and conclusions}

There are lots of algorithms to analyze clustering in networks, bipartite or otherwise, so why another one? The algorithm we presented here has certain important advantages with respect to previous ones. The main one is its faster performance. It is based on two operations: first of all, conducting a FET between every pair of entities, and secondly, running SLINK to obtain a hierarchical clustering of the entities based on the outcomes of the pairwise FETs. Both operations have a complexity $O(n^2)$ when applied to a bipartite network of $n$ entities---or to a one-mode network with $n$ nodes. The fastest modularity-based algorithm requires a matrix diagonalization, whose complexity is $O(n^3)$.

But this is not the only advantage of our algorithm: its outcome is a multi-resolution analysis of the relations between the nodes in the form of a dendrogram. If so needed, we can make use of the susceptibility measure that we have introduced to determine an `optimal' partition of the nodes. This measure has the bonus of quantifying the quality of the partition (the higher the susceptibility the better the partition). But we should not neglect that the multi-resolution clustering provided by the dendrogram contains useful information that remains hidden in standard clustering algorithms---see for example our analyses of survey data in section \ref{sec:surveys} and how they compare to standard techniques such as MCA (Sec.~\ref{sec:surveys}).

Furthermore, typically there are nodes that do not fit in any cluster of the dendrogram. The fact that modularity-based algorithms always assign every node to a cluster may mislead us to think of this issue as a drawback of the method. On the contrary, as the example of the congressional voting records shows (Sec.~\ref{sec:voting}), it may be revealing a special status of these nodes (in case of voting, the relative independence of some congressmen). This information is often as useful as that of the clustering itself.

Finally, the dissimilarity introduced by the FET is statistically meaningful: it measures the probability that observing those coincidences between the features of two entities is purely due to chance, and this information is normally lacking in other clustering algorithms.

The availability of a dendrogram prevents some problems inherent to the optimization of a network measure such as modularity. It has been shown \cite{guimera:2004,radicchi:2010} that modularity-based methods can find spurious structure in random networks. Fluctuations are a source of meaningless associations, but as we have shown, they are easy to spot on a dendrogram. To begin with, there are no obvious clusters that partition the network in big blocks, and furthermore, the $p$-values for which associations occur are too high to be statistically meaningful. Of course, the calculation of the susceptibility will always provide an `optimal' partition even for a random network, however its small value is an indication that this clustering is not to be trusted.

Our algorithm is very easy to implement as well, given the availability of efficient algorithms for calculating the $p$-value of a FET and for performing the hierarchical clustering. As a matter of fact, we provide an open-access, documented implementation of the complete algorithm in Python, \emph{clusterBip}, for public download. The code is available on GitHub \url{https://github.com/mpereda/clusterBip}.

\section*{Acknowledgments}


This research has been funded by the Spanish Ministerio de Ciencia, Innovaci\'on y Uni\-ver\-si\-da\-des-FEDER funds of the European Union support, under project BASIC (PGC2018-098186-B-I00).

\bibliography{bipartite}

\end{document}